\documentclass[12pt,onecolumn]{IEEEtran}

\usepackage{algorithm}
\usepackage{algorithmic}

\usepackage[psamsfonts]{amsfonts}
\usepackage{amsmath}
\usepackage{eufrak}
\usepackage{amssymb}
\usepackage{graphics}
\usepackage{psfrag}
\usepackage{epsfig}
\usepackage{subfigure}
\usepackage{array}
\usepackage{multirow}

\IEEEoverridecommandlockouts

\title{Efficient Symbol Sorting for High Intermediate Recovery Rate of LT Codes}
\author{Ali Talari, Behzad Shahrasbi, and Nazanin Rahnavard\\
Oklahoma State University, Stillwater, OK 74078 \\
Emails: {\{ali.talari, behzad.shahrasbi, nazanin.rahnavard\}@okstate.edu}}

\begin{document}
\maketitle

\begin{abstract}

LT codes are modern and efficient \emph{rateless forward error correction} (FEC) codes with close to channel capacity performance. Nevertheless, in \emph{intermediate range} where the number of received encoded symbols is less than the number of source symbols, LT codes have very low recovery rates.

In this paper, we propose a novel algorithm which significantly increases the intermediate recovery rate of LT codes, while it preserves the codes' close to channel capacity performance. To increase the intermediate recovery rate, our proposed algorithm \emph{rearranges} the transmission order of the encoded symbols exploiting their structure, their transmission history, and an estimate of the channel's erasure rate. We implement our algorithm for conventional LT codes, and numerically evaluate its performance.
\end{abstract}


\section{Introduction}


\emph{LT codes} are modern, efficient, and flexible \emph{rateless forward error correction} (FEC) codes \cite{Luby02lt}. An LT encoder at a source $S$ can potentially generate a limitless number of encoded symbols $c_i,i\in\{0,1,\ldots\}$ from $k$ source symbols $\underline{x} = \{x_1, x_2,\ldots,x_k\}$. The decoder at a destination $D$ can successfully recover $\underline{x}$ from any $k\gamma$ of received encoded symbols, where $\gamma$ is the received \emph{coding overhead}. Let $\gamma_{succ}$ denote the required overhead for a successful decoding. Conventional LT codes can achieve $z\approx1$ with high probability at $\gamma_{succ}$'s slightly larger than one, where $z$ denotes the ratio of the number of recovered source symbols to $k$ at $D$. Note that, for LT codes, $\gamma_{succ}$ is fixed and known \cite{Luby02lt}.



Each LT code is identified by a \emph{degree distribution}. In LT encoding, first an encoded symbol degree $d$ is chosen from a degree distribution, $\{\Omega_{1},\Omega_{2},\ldots,\Omega_{k}\}$, where $\Omega_{i}$ is the probability that $d=i$. This degree distribution is also shown by its generator polynomial $\Omega(y)=\sum_{i=1}^{k}\Omega_{i}y^{i}$. Next, $d$ $x_j$'s are chosen uniformly at random from $\underline{x}$, and are \emph{XOR}ed to generate an encoded symbol $c_i$. Let $\varepsilon\in [0,1)$ denote the channel erasure rate. Therefore, $S$ needs to generate on average $m=\frac{k\gamma_{succ}}{1-\varepsilon}$ encoded symbols so that $D$ can collect $k\gamma_{succ}$ of them for a successful decoding.


In LT decoding at $D$, if a newly delivered $c_i, i \in\{1,2,\ldots,m\}$ has degree-one, it can decode one $x_j,j\in\{1,2,\ldots,k\}$. If the degree of $c_i$ is larger than one then previously recovered $x_j$'s, if any, are removed from $c_i$ to decrease its degree. If the degree of $c_i$ is reduced to one, this symbol is similar to a new degree-one $c_i$. On the other hand, if the remaining degree is larger than one, the symbol is \emph{buffered}. If any new $x_j$ is recovered in the previous step, it is removed from the buffered $c_i$'s. Hence, some buffered $c_i$'s may acquire a degree equal to one, resulting in recovery of further $x_j$'s. This procedure is repeated iteratively until no more degree-one $c_i$ emerges.

\emph{Intermediate range} of LT codes refers to the case where the transmission is still in progress, i.e., $0\leq\gamma<1$ \cite{Sanghavi07intermeidate}. Since conventional LT codes \cite{Luby02lt} are designed to have an almost complete recovery of $\underline{x}$ ($z\approx1$) for certain $\gamma_{succ}\text{'s}>1$, they have low intermediate recovery rates ($z \approx 0$) at $0\leq\gamma<1$. However, in many applications where \emph{partial} recovery of the source symbols from the incomplete received encoded symbols is still beneficial, intermediate recovery rate becomes important. For instance, in multimedia transmission the receiver can play a lower quality of the multimedia contents from the incomplete recovered data. This motivates the design of an algorithm to improve the intermediate recovery rate of existing LT codes.

In order to obtain a high $z$ (as close as possible to $\gamma$) in $0\leq\gamma<1$, each delivered $c_i$ should decode (on average) one $x_j$ instead of being buffered. We propose an algorithm to \emph{rearrange} the transmission order of $c_i$'s based on their transmission history, structure of each $c_i$, and an estimate of $\varepsilon$, such that each delivered $c_i$ can decode one $x_j$ with high probability. By employing our proposed algorithm, the intermediate performance of a given LT code improves significantly, while the recovery rate of $\underline{x}$ at $\gamma_{succ}$ remains intact. In other words, the code remains capacity achieving similar to its original setup.


This paper is organized as follows. Section \ref{related_work} reviews the exiting work on intermediate performance of LT codes. In Section \ref{algorithm}, we propose our novel sorting algorithm. Next, Section \ref{evaluation} reports the performance evaluation of our proposed algorithm. Finally, Section \ref{conclusion} concludes the paper.



\section{Related Work} \label{related_work}


In \cite{Sanghavi07intermeidate}, Sanghavi has shown that $0\leq\gamma<1$ and equivalently $0\leq z<1$ can be divided into three regions of $z \in [0,\frac{1}{2}]$, $z \in [\frac{1}{2},\frac{2}{3}]$, and $z \in (\frac{2}{3},1)$. In each region, the upper bound on $z$ for all rateless codes versus $\gamma$ is formulated, and the optimum $\Omega(y)$'s to gain these upper bounds are provided. The $\Omega(y)$'s provided for each region perform optimally in that specific region only, thus they have a low $z$ compared to $\gamma$ in other two regions. Further, provided $\Omega(y)$'s are not capacity achieving, and they are designed for asymptotic case (infinite $k$) and may not be employed in practice where $k$ is finite.

Authors in \cite{Kamra06growthcodes} have proposed \emph{Growth codes}, which are designed to increase the number of recovered $x_j$'s in intermediate range in wireless sensor networks. In Growth coding, $S$ gradually increases the degree of $c_i$'s on-the-fly based on the value of $z$ (which is known to $S$ by feedbacks received from $D$), such that the instantaneous decoding probability of each delivered $c_i$ is maximized. Growth codes only consider the instantaneous recovery probability of each $c_i$, thus they do not have a close to capacity performance. More importantly, Growth codes require a lot of feedbacks from the receiver.

Authors in \cite{Kim09improved} have proposed to employ multiple feedbacks to transmit $c_i$'s in the order of their \emph{degree} to increase the intermediate recovery rate. Since the decoding of $c_i$'s with lower degrees depends on the recovery of a smaller subset of $x_j$'s, they have a higher probability of decoding an $x_j$ in $D$ at the beginning of transmission. Consequently, an improvement is observed in the intermediate recovery rate of LT codes. However, the algorithm proposed in \cite{Kim09improved} cannot outperform the code of \cite{Kamra06growthcodes, Sanghavi07intermeidate} in intermediate range. Besides, we show that our proposed algorithm always surpasses the algorithm in \cite{Kim09improved}.

In our recent work \cite{talari09intermediate}, we have designed several $\Omega(y)$'s for LT codes to obtain optimum intermediate performance \emph{throughout} the intermediate range rather than a single $\gamma$ \cite{Sanghavi07intermeidate}. The LT codes designed in \cite{talari09intermediate} do not require channel information or feedbacks in contrast to \cite{Kamra06growthcodes, Kim09improved}. However, similar to \cite{Sanghavi07intermeidate} the codes designed in \cite{talari09intermediate} cannot achieve channel capacity.


\section{The Proposed Algorithm} \label{algorithm}


\subsection{Discussion and Idea}


In the previous section, we observed a \emph{trade-off} between being channel capacity achieving and having a high intermediate recovery rate. Some LT codes have optimum intermediate performance but they cannot achieve channel capacity \cite{talari09intermediate, Sanghavi07intermeidate}. On the other hand, the algorithms that are proposed to improve the intermediate recovery rate of LT codes with close to channel capacity such as \cite{Kim09improved} cannot outperform codes of the first group in intermediate range.

The reason for this trade-off is that the codes in the first group have $\Omega(y)$'s that result in generation of a large fraction of low-degree $c_i$'s. Low-degree $c_i$'s have a higher probability of decoding a source symbol when $z$ is small. However as $z$ grows, some of the received $c_i$'s become redundant and cannot recover any $x_j$ due to earlier recovery of all their adjacent $x_j$'s. Therefore, these codes cannot have a close to channel capacity performance.

On the other hand, $\Omega(y)$'s of codes with close to channel capacity performance result in generation of $c_i$'s with much higher degrees. Decoding of high-degree $c_i$'s depends on the recovery of many $x_i$'s at $D$, thus in intermediate range these $c_i$'s are mostly buffered for a later decoding. Therefore, for these codes $z$ does not grow considerably with $\gamma$. However, the buffered high-degree $c_i$'s are simultaneously decoded together and recover $\underline{x}$ at $\gamma_{succ}$, which makes the code capacity achieving.

We can see that in LT codes with capacity achieving performance, each $c_i$ eventually decodes close to one $x_j$ on average, since all $k$ $x_j$'s are decoded from $k \gamma_{succ}$ $c_i$'s, which are slightly more than $k$ symbols. Consequently, if $c_i$'s are transmitted in the \emph{order} that they are \emph{decoded}, we can significantly improve the intermediate recovery rate of \emph{capacity achieving} LT codes. We propose our algorithm to transmit $c_i$'s in this \emph{correct order} for two cases of constant and varying $\varepsilon$.



\subsection{Algorithm for Constant $\varepsilon$}

Similar to \cite{Gummadi08relaying}, we assume that an estimate of the channel erasure rate, $\varepsilon$, is available at $S$. Since we assume that $D$ generates no feedbacks, our algorithm is designed to be implemented on the encoder side. Therefore, our algorithm can exploit the information available in $S$ only, and the decoder remains intact.

In conventional LT coded symbol transmission, $S$ generates $m=\frac{k\gamma_{succ}}{1-\varepsilon}$ random $c_i$'s from a capacity achieving degree distribution $\Omega(.)$ such as \emph{Robust-Soliton} degree distribution \cite{Luby02lt}. After transmission, $D$ receives $k\gamma_{succ}$ encoded symbols, which results in a successful decoding of $\underline{x}$. This method has a poor intermediate recovery rate as we later show.

In our proposed scheme, after generating $m$ $c_i$'s, $S$ performs as follows. $S$ maintains a probability vector $\underline{\rho}=[\rho(1),\rho(2),\ldots,\rho(k)]$, in which $\rho(j)$ represents the probability that $x_j$ is still \emph{not recovered at} $D$. Clearly, $S$ sets $\underline{\rho}$ to an all-one vector when the transmission has not started yet since no $x_j$ is recovered at $D$. At each transmission and based on $\underline{\rho}$, $S$ finds a $c_i$ that has the highest probability of recovering an $x_i$ at $D$ (as we later describe in Algorithm \ref{sorting_algth}). Next, $S$ transmits $c_i$ and updates $\rho(j), j\in \mathcal{N}(c_i)$, where $\mathcal{N}(c_i)$ represents the set of indices of $x_i$'s \emph{XOR}ed together to generate $c_i$. $S$ continues until all $m$ $c_i$'s are transmitted.

An encoded symbol $c_i$ with degree $d$, i.e., $|\mathcal{N}(c_i)|=d$, where $|.|$ represents the cardinality of a set, can recover a source symbol $x_j$ iff all $x_k,k \in \{\mathcal{N}(c_i)-j\}$ has already been recovered. Let $\underline{p_{dec}}=\{p_{dec}(1), p_{dec}(2),\ldots,p_{dec}(m)\}$, where $p_{dec}(i)$ is the probability that $c_i$ can recover a source symbol at $D$, and $p_{dec}(i)=0$ if $c_i$ has been previously transmitted. Since at the beginning of transmission no source symbol is still recovered, we have $p_{dec}(i)=0$ if $|\mathcal{N}(c_i)|>1$, i.e., $c_i$'s with degrees larger than one cannot decode any $x_j$ at $D$. Besides, for $|\mathcal{N}(c_i)|=1$ we have $p_{dec}(i) = (1-\varepsilon)$, i.e. only degree-one encoded symbols that are not erased by the channel (with probability $(1-\varepsilon)$) can recover a source symbol.

We can see that at the beginning of transmission degree-one $c_i$'s have the highest probabilities of decoding an $x_j$ at $D$. Therefore, $S$ transmits degree-one $c_i$'s with $\mathcal{N}(c_i)=\{j\}$, and updates $\rho(j)= \varepsilon \rho_{old}(j)$, where $\rho_{old}(j)$ is the value of $\rho(j)$ before $c_i$ was transmitted.



%






Next, we consider a degree-two $c_i, \mathcal{N}(c_i)=\{j,k\}$. $c_i$ can recover $x_j$ with probability $(1-\varepsilon) (1-\rho(j))\rho(k)$, which is the probability that $c_i$ is not dropped on channel, $x_j$ has not been recovered previously, and the $x_k$ has already been recovered. Similarly, $c_i$ can recover $x_k$ with probability $(1-\varepsilon) (1-\rho(k))\rho(j)$. Consequently, $p_{dec}(i) = (1-\varepsilon) [(1-\rho(k))\rho(j)+(1-\rho(j))\rho(k)]$. Assume $\forall l\not =i, p_{dec}(i) > p_{dec}(l)$, i.e. $c_i$ has the highest probability of decoding an $x_j$ at $D$. Therefore, $S$ transmits $c_i$ next and sets $\rho(j)= \rho_{old}(j) (1-(1-\varepsilon) (1-\rho_{old}(k)))$ and $\rho(k)= \rho_{old}(k) (1-(1-\varepsilon) (1-\rho_{old}(j)))$.


Further, we consider $c_i, |\mathcal{N}(c_i)|=d$. If $c_i$ is successfully delivered to $D$, it can possibly decode $x_j, j \in \mathcal{N}(c_i)$. Similar to low degree $c_i$'s, $x_j$ can be decoded with probability $(1-\varepsilon) \rho(j) \prod \limits_{v \in \mathcal{N}(c_i), v \not = j}(1-\rho(v))$. Therefore, $p_{dec}(i) = (1-\varepsilon) \sum \limits_{l \in \mathcal{N}(c_i)} [\rho(l) \prod \limits_{v \in \mathcal{N}(c_i),v \not = l}(1-\rho(v))]$. If $p_{dec}(i)>p_{dec}(l),\forall l\not =i$, $S$ transmits $c_i$ and updates $\rho(j)=\rho_{old}(j)[1-(1-\varepsilon)\prod \limits_{v \in \mathcal{N}(c_i),v \not = j}(1-\rho_{old}(v))], j \in \mathcal{N}(c_i)$.



We summarize our proposed sorting scheme in Algorithm \ref{sorting_algth}. The output of Algorithm \ref{sorting_algth} is the suitable rearranged transmission order $\underline{\pi}$ of $c_i$'s that substantially improves $z$ in $0\leq\gamma<1$. In this algorithm, $argmax(\underline{p_{dec}})$ is a function that returns $i$ where $\forall j\not = i, p_{dec}(i) > p_{dec}(j)$. Further, if $c_i$ and $c_l$ both have the highest probability of decoding of an $x_j$, i.e., $p_{dec}(l) = p_{dec}(i)$, then $argmax(\underline{p_{dec}})$ returns the index of $c_i$ or $c_l$, whichever has the \emph{lowest} degree. Further, if $c_i$ and $c_l$ have equal degrees (similar to degree one $c_i$'s at the beginning of transmission), $argmax(\underline{p_{dec}})$ randomly returns one of the indices. Clearly, we are assigning earlier transmission priority to lower degree symbols, which we later show is important for $\varepsilon\rightarrow 1$.

\begin{algorithm}
\caption{The proposed symbol sorting algorithm}
\label{sorting_algth}
\begin{algorithmic}
\STATE Initialize: $\underline{\pi}=\emptyset$, $\underline{\rho}=\{1\}_{1\times k}$
\WHILE{$|\underline{\pi}|<m$} 
    \FOR{$j = 1$ to $m, j \not\in \underline{\pi}$}
        \STATE $p_{dec}(j)=(1-\varepsilon)\sum \limits_{l \in \mathcal{N}(c_i)} [\rho(l) \prod \limits_{v \in \mathcal{N}(c_i),v \not = l}(1-\rho(v))]$
    \ENDFOR
    \STATE $i^* = argmax(\underline{p_{dec}})$
    \STATE  $\underline{\pi} =[\underline{\pi} , i^*]$
    \FOR{$j \in \mathcal{N}(c_{i^*})$}
        \STATE $\rho(j)=\rho_{old}(j)[1-(1-\varepsilon)\prod \limits_{v \in \mathcal{N}(c_{i^*}),v \not = j}(1-\rho_{old}(v))]$
    \ENDFOR
\ENDWHILE
\end{algorithmic}
\end{algorithm}




Our proposed algorithm increases the coding complexity of LT codes from $O(k\log k)$ \cite{Luby02lt} to $O(k^2)$, while it does not deteriorate the decoding complexity.

Further, in our proposed algorithm all $c_i$'s need to be generated and sorted before the transmission can start in contrast to the conventional LT coding where each $c_i$ can be independently transmitted upon generation. Therefore, some delays may be introduced.

However, this delay can be easily eliminated with the following procedure. Clearly, the performance of our proposed scheme is independent of $\underline{x}$'s contents and only depends on $\mathcal{N}(c_i),i\in\{1,2,\ldots,m\}$ and $\underline{\pi}$. Therefore, before the transmission starts, $S$ generates $c_i$'s from a dummy $\underline{x}$, and obtains an off-line version of $\underline{\pi}_{\text{off-line}}$. $S$ saves $\mathcal{N}_{\text{off-line}}(c_i)$, and $\underline{\pi}_{\text{off-line}}$ for a later use. When the actual encoding starts, $\underline{x}$ of interest replaces the dummy $\underline{x}$, and $S$ generates $m$ $c_i$'s \emph{in the order} dictated by  $\underline{\pi}_{\text{off-line}}$, \emph{XOR}ing $x_j , j\in\ \mathcal{N}_{\text{off-line}}(c_i)$. In this way, each $c_i$ can be transmitted upon generation and the delay is completely eliminated at the cost of some data storage.


In the next section, we extend Algorithm \ref{sorting_algth} to the case where $\varepsilon$ varies.


\subsection{Algorithm for Varying $\varepsilon$}

Assume that $S$ has generated $m$ $c_i$'s considering $\varepsilon$. Assume that the erasure rate of the channel changes to $\varepsilon_{new}$ when $\frac{k\gamma_{c}}{1-\varepsilon}$ symbols has already been transmitted so that $\frac{k(\gamma_{succ}-\gamma_{c})}{1-\varepsilon}$ $c_i$'s are still remaining to be transmitted.

If  $\varepsilon_{new} > \varepsilon$, less than $k\gamma_{succ}$ $c_i$'s would be collected by $D$, making the full decoding impossible. In this case, $S$ generates $t=(\frac{1}{1-\varepsilon_{new}}-\frac{1}{1-\varepsilon})k(\gamma_{succ}-\gamma_{c})$ \emph{new} $c_i$'s, and adds them to the queue of $c_i$'s to be transmitted to ensure the delivery of $k\gamma_{succ}$ $c_i$'s to $D$. Next, $S$ \emph{rearranges} all $c_i$'s in the queue according to Algorithm \ref{sorting_algth} and continues the transmission.


In the second case for $\varepsilon_{new} < \varepsilon$, $S$ randomly drops $1-\frac{1-\varepsilon}{1-\varepsilon_{new}}$ fraction of remaining $c_i$'s from the transmission queue. This limits the number of delivered $c_i$'s to $k\gamma_{succ}$, hence the code maintains its close to channel capacity performance.

If the erasure rate of the channel varies several times, the same procedures are followed after each change. We assume that $\varepsilon_{new}$ is known to $S$, which can be realized by receiving few feedbacks from the receiver.

Note that the symbol dropping procedure described above is similar to \emph{puncturing LDPC} \cite{Pishro-Nik05Nonuniform} and \emph{turbo} codes \cite{Douglas97Rate-compatible} to achieve a certain higher coding rate for these codes.


\vspace{-1mm}

\section{Evaluation of the Proposed Algorithm} \label{evaluation}

\vspace{-1mm}

We emphasize that our proposed algorithm can be applied to an LT code with any degree distribution to increase its intermediate recovery rate when $\varepsilon$ is available at $S$. The advantage of this algorithm is that if the code is capacity achieving, it remains capacity achieving after employing our algorithm.

To evaluate the performance of our proposed algorithm, we implement it for two well-known LT codes. The first code we employ is the LT code used in \emph{Raptor codes} \cite{Shokrollahi06raptor} with degree distribution $\Omega(y)$ given below and $k=1000$. Since low error floors cannot be achieved in intermediate range, the precoding phase of Raptor codes can be skipped.

\vspace{-3mm}

\begin{equation*}
\footnotesize{
\begin{split}
\Omega(y)=0.00797x+0.49357x^{2}+0.16622x^{3}+0.07265x^{4}+0.08256x^{5}\\
             +0.05606x^{8}+0.03723x^{9}+0.05559x^{19} + 0.02502x^{65}+0.00314x^{66}.
\end{split}
}
\end{equation*}

The second code is an LT code with $k=1000$ and Robust-Soliton degree distribution \cite{Luby02lt} with parameters $c=0.05$ and $\delta=0.01$.

\begin{figure}[h]
\includegraphics[width=3.4in,angle=0]{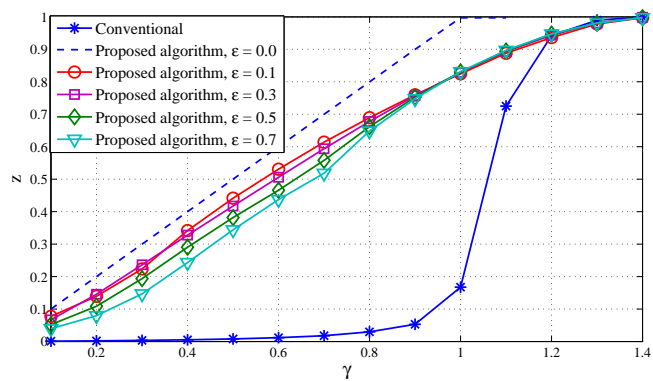}
\caption{Improvement made in the intermediate recovery rate of a Raptor code proposed in \cite{Shokrollahi06raptor} employing proposed algorithms for various $\varepsilon$'s versus $\gamma$.}
\label{result_shok}
\end{figure}

\begin{figure}[h]
\includegraphics[width=3.4in,angle=0]{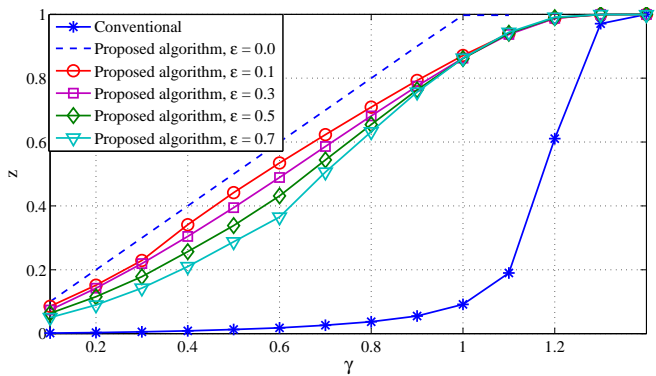}
\caption{Improvement made in the intermediate recovery rate of an LT code \cite{Luby02lt} employing proposed algorithms for various $\varepsilon$'s versus $\gamma$.}
\label{result_rsd}
\end{figure}

Figures \ref{result_shok} and \ref{result_rsd} illustrate the improvement made in $z$ for aforementioned Raptor and LT codes employing our proposed sorting algorithm.

From Figures \ref{result_shok} and \ref{result_rsd}, we can see a significant improvement in $z$ for both codes. For instance at $\gamma=1$, for the first and the second code we can see $382\%$ and $843\%$ improvement in $z$, respectively. We can observe that the amount improvement depends on the value of $\varepsilon$. The rationale behind this is that when $\varepsilon$ decreases our algorithm can estimate the recovery probability of $x_j$'s more accurately, which results in a more efficient reordering of $c_i$'s. As $\varepsilon$ becomes larger, the ordering of $c_i$'s becomes less accurate. It is worth noting that the conventional transmission of LT codes results in the same curve of $z$ regardless of $\varepsilon$'s value.

\subsection{Upper and Lower Bounds on Algorithm's Performance}

As $\varepsilon\rightarrow 1$, $S$ cannot make a meaningful estimation about the recovery of $x_j$'s at $D$, and $\underline{\rho}$ always remains an all-one vector. Since in our proposed algorithm, $c_i$'s with lower degrees have higher priority of transmission, for $\varepsilon\rightarrow 1$ our algorithm approaches to the case where $c_i$'s are only sorted based on their degrees in $\underline{\pi}$. Consequently, in this case our proposed algorithm boils down to the algorithm proposed in \cite{Kim09improved}. As a result, the improvement made by our algorithm is lower bounded by the results of \cite{Kim09improved}. Moreover, for $\varepsilon\rightarrow 0$, $S$ can estimate which $x_j$'s are being decoded with a high accuracy, thus more exact $\underline{\pi}$ can be acquired, and the intermediate performance approaches the ideal upper bound, i.e., $z=\gamma$.

\begin{figure}[h]
\includegraphics[width=3.4in,angle=0]{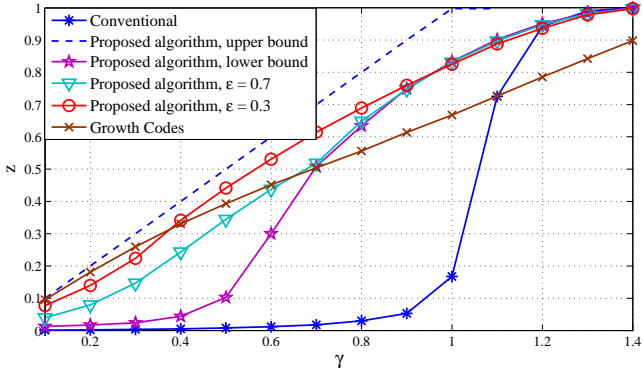}
\caption{The upper and the lower bounds on the improvement made by our proposed scheme for Raptor codes \cite{Shokrollahi06raptor}, compared to the intermediate recovery rate of Growth codes and conventional LT codes.}
\label{result_shok_bounds}
\end{figure}

\begin{figure}[h]
\includegraphics[width=3.4in,angle=0]{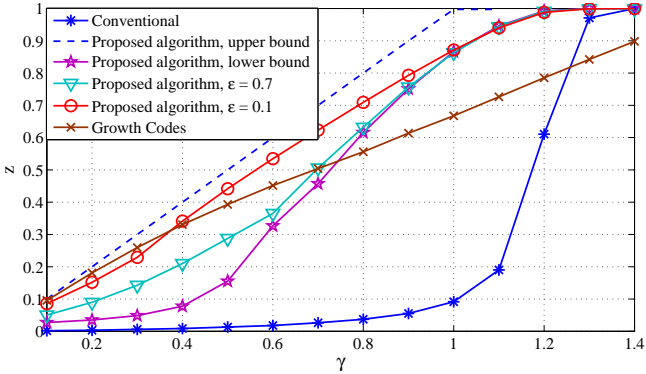}
\caption{The upper and the lower bounds on the improvement made by our proposed scheme for LT codes \cite{Luby02lt}, compared to the intermediate recovery rate of Growth codes and conventional LT codes.}
\label{result_rsd_bounds}
\end{figure}

The upper and the lower bounds on our proposed scheme are depicted in Figures \ref{result_shok_bounds} and \ref{result_rsd_bounds} for distribution of Raptor and LT codes, respectively. For comparison, we have also provided the recovery rate curves of Growth codes \cite{Kamra06growthcodes} and conventional LT transmission for the same codes. Further, the lower bound illustrates the performance of the scheme proposed in \cite{Kim09improved}.



Figures \ref{result_shok_bounds} and \ref{result_rsd_bounds} show considerable improvement for intermediate recovery rate of both employed LT codes for any value of $\varepsilon$. We can see that Growth codes outperform our propose algorithm only for a small region of $\gamma$, while they cannot have a close to channel capacity performance.


\subsection{Algorithm's Performance for Varying $\varepsilon$}



\begin{figure}[h]
\includegraphics[width=3.2in,angle=0]{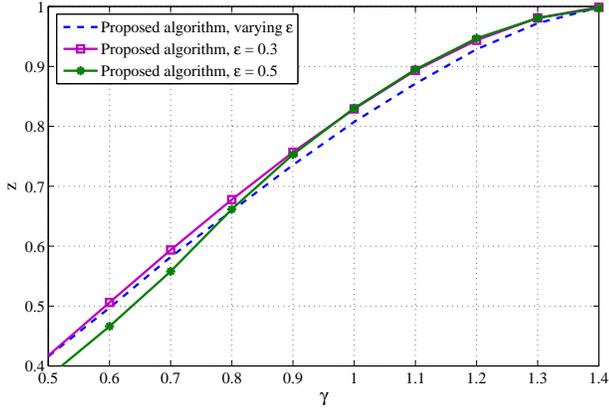}
\caption{The intermediate performance of a Raptor code employing our proposed algorithm for the case where $\varepsilon$ increases from $0.3$ to $0.5$ with $\gamma_c=0.5$.}
\label{varying}
\end{figure}

As described earlier, if $\varepsilon$ increases, $S$ needs to generate some new $c_i$'s. Assume that $S$ is transmitting $m$ Raptor encoded $c_i$'s (without precoding) generated for $\varepsilon=0.3$. Also assume that $\varepsilon$ increases to $\varepsilon_{new}=0.5$ at $\gamma_c = 0.5$. Based on our proposed algorithm, $S$ adds $t = \lceil 0.5714 k (\gamma_{succ}-\gamma_c)\rceil$ new $c_i$'s to the queue, and updates $\underline{\pi}$ accordingly. We have depicted $z$ versus $\gamma$ for this case in Figure \ref{varying}. For comparison, we have also depicted $z$ for constant $\varepsilon\in\{0.3,0.5\}$.

Figure \ref{varying} shows that fluctuation in the $\varepsilon$ is well compensated by our algorithm. Since the newly generated $c_i$'s at $\gamma_c$ disturb the original sorting order, a slight degradation in $z$ can also be observed. These new $c_i$'s might have been transmitted earlier than $\gamma_c$ if they were present in the queue from the beginning. In spite of all the fluctuation in intermediate performance, the code remains capacity achieving since $D$ can collect $k\gamma_{succ}$ encoded symbols in total.

\subsection{Comparison with Fixed-Rate Codes}

Since we assumed that an estimate of the channel loss rate, $\varepsilon$, is available at the source, $S$ can employ existing \emph{fixed-rate} codes instead of LT codes to encoded $\underline{x}$. Therefore, we need to compare the performance of our proposed scheme with the intermediate recovery rate of existing fixed-rate codes.

\begin{figure}[h]
\includegraphics[width=3.2in,angle=0]{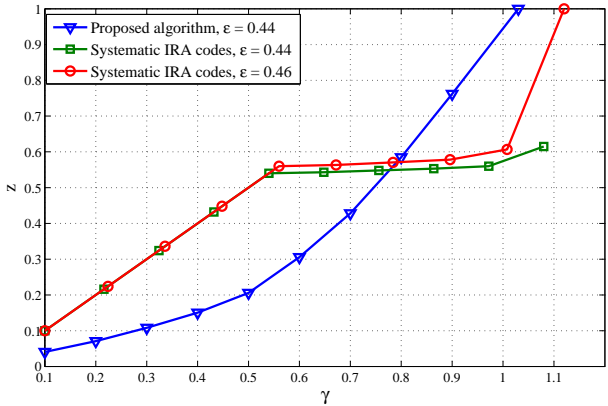}
\caption{The intermediate recovery rate of systematic IRA codes \cite{jin00irregular} compared to our proposed algorithm employing Raptor codes \cite{Shokrollahi06raptor} for $k=10000$.}
\label{IRA_comparison}
\end{figure}

\emph{Systematic irregular repeat-accumulate (Systematic IRA)} codes \cite{jin00irregular} are capacity achieving fixed-rate codes on erasure channels, which can provide a high intermediate recovery rate compared to other existing fixed-rate codes. Figure \ref{IRA_comparison} shows the intermediate recovery rate of a systematic IRA code with rate $R=0.5$ versus our proposed algorithm employing Raptor codes without precoding for two cases of $\varepsilon=0.44$ and $\varepsilon=0.46$ with $k=10000$.


\vspace{-3mm}

\vspace{-2mm}

From Figure \ref{IRA_comparison}, we can see that the employed systematic IRA code outperforms our proposed scheme for small values of $\gamma$. This high intermediate recovery is due to the systematic part of these codes which results in transmission of uncoded $\underline{x}$. When the systematic part of the code is being transmitted, each delivered $c_i$ is itself a source symbol, hence $z$ and $\gamma$ are equal. However, we can observe that when the transmission of systematic part ends, $z$ does not increase linearly with $\gamma$ anymore and our proposed scheme outperforms this code for larger values of $\gamma$. Further, as $\varepsilon\rightarrow 1$ the systematic part cannot be delivered and the gain from the systematic part is eliminated. In this case, our proposed algorithm always outperforms systematic IRA codes.

Furthermore, fixed-rate codes seriously suffer from their fixed rates since they cannot compensate slight variations in $\varepsilon$. For instance, we can see that when $\varepsilon$ is increased from $0.44$ to $0.46$, $z$ decreases from $1$ to $0.6150$ at the end of transmission. However, as observed in Figure \ref{varying} our algorithm employed along with a Raptor code exhibit good performance in spite of very large variations in $\varepsilon$.


\vspace{-1mm}

\section{Conclusion}\label{conclusion}

In this paper, we proposed an algorithm to increase the intermediate recovery rate of capacity achieving LT codes. In our proposed algorithm, the transmitter exploits the structure of LT encoded symbols, a history of previously transmitted encoded symbols, and an estimate of channel's erasure rate to sort the transmission order of the encoded symbols to gain a high intermediate recovery rate.

Using numerical results, we showed that our algorithm can increase the intermediate recovery rate of LT codes to a great extent while the code remains capacity achieving. We also showed that this algorithm performs well for fluctuating channel erasure rates.


To extend this work, we intend to implement the proposed algorithm for real multimedia transmission and observe the improvement made in the quality of the received stream.




\vspace{-1mm}

\bibliographystyle{ieeetr}
\bibliography{ref}

\end{document}